\documentclass[twocolumn,aps,pre,floatfix,superscriptaddress]{revtex4-1}
\usepackage{graphicx}
\usepackage{epic,eepic}
\usepackage{graphics}
\usepackage{epsfig}
\usepackage{subfigure}

\newcommand{\be}{\begin{equation}}
\newcommand{\ee}{\end{equation}}
\newcommand{\bea}{\begin{eqnarray}}
\newcommand{\eea}{\end{eqnarray}}

\usepackage{amsmath}
\usepackage{amsfonts}
\usepackage{amssymb}
\usepackage{bm}
\usepackage{color}
\usepackage{graphicx}
\usepackage[utf8x]{inputenc}

\begin{document}
\title{Hydrodynamic and geometric effects in the sedimentation of model run-and-tumble bacteria}

\author{A. Scagliarini}
\email{andrea.scagliarini@cnr.it}
\affiliation{IAC-CNR, Istituto per le Applicazioni del Calcolo ``Mauro Picone'', Via dei Taurini 19, 00185 Rome, Italy.}%
\affiliation{INFN, sezione Roma ``Tor Vergata'', via della Ricerca Scientifica 1, 00133 Rome, Italy}
\author{I. Pagonabarraga}
\affiliation{CECAM, Centre Europ\'een de Calcul Atomique et Mol\'eculaire, Ecole Polytechnique  F\'ed\'erale  de  Lausanne,  Batochimie, Avenue Forel 2, 1015 Lausanne, Switzerland}
\affiliation{Departament de F\'{\i}sica de la Mat\`eria Condensada, Universitat de Barcelona, Mart\'{\i} i Franqu\`es 1, 08028 Barcelona, Spain}
\affiliation{Universitat de Barcelona, Institute of Complex Systems (UBICS), Universitat de Barcelona, 08028 Barcelona, Spain}

\begin{abstract}
The sedimentation process in a suspension of bacteria is the result 
of the competition between 
gravity and the intrinsic motion of the microorganisms. We perform simulations of
run-and-tumble ``squirmers'' that move in a fluid medium, focusing on the dependence of the 
non-equilibrium steady state on the bacterial swimming properties.
We find that for high enough activity, the density profiles are no longer simple exponentials;
we recover the numerical results via the introduction of a local effective temperature,
suggesting that the breakdown of the Perrin-like exponential form is a 
collective effect due to the onset of fluid-mediated dynamic correlations among particles.
We show that analogous concepts can fit also the case of shakers, for which we report the
 first study of this kind. Moreover we provide evidences of scenarios
 where the solvent hydrodynamics induces non-local effects which require
 the fully three-dimensional dynamics to be taken into account in order to
 understand sedimentation of active suspensions.   
Finally, analyzing the statistics of the bacterial swimming orientations, we discuss the
emergence of polar order in the steady state sedimentation profiles.
\end{abstract}

\maketitle

\section{Introduction}
A number of microorganisms (bacteria, algae, etc...) have the ability to swim 
in a liquid environment through the generation of autonomous motion at expenses of their metabolism, 
thus being intrinsically out-of-equilibrium. 
As such, these systems lead to new challenges such as the understanding of how 
collective phenomena and self-organization emerge from the relevant features of their propulsion
mechanism~\cite{Marchetti,Elgeti,Simha,Llopis}.  
In this perspective a suspension of active particles is qualitatively different from a suspension of 
passive ones. Maybe the simplest, yet not trivial, example of this is the case of 
a constant external forcing on the suspension, such as gravity in the sedimentation process.
In fact, when thermal fluctuations are negligible (as it is in the case of particles above the micron size), while
passive particles would inevitably precipitate, active suspensions maintain a finite sedimentation length that grows
with the self-propulsion speed. This result was predicted theoretically for ``dry'' suspensions (i.e. where the solvent 
hydrodynamics in neglected) of non-interacting run-and-tumble particles~\cite{Tailleur1,Tailleur2} and, then, confirmed
in numerical simulations with point-like dipoles~\cite{Nash} and experimentally in suspensions of active colloids~\cite{Palacci}.
Suspensions of self-propelled particles under gravity have been also reported to display a non-trivial orientational
dynamics, with the development of an associated polar order~\cite{Enculescu,Kuhr} or even,
in the case of bottom-heavy particles, to the inversion
of the sedimentation profiles~\cite{Wolff}. 
In this paper we present a computational study of bacterial sedimentation, 
where hydrodynamics is fully resolved near and far from the swimmer's
surface. We provide evidence that hydrodynamic
correlations induce important deviations form the phenomenology for dry suspensions
in the steady state of both self-propelled swimmers and ``shakers'' (for
which, to the best of our knowledge, this represents the first study of this kind).
The sedimentation profiles observed when bacterial activity is
intense are captured
through a simple extension of a drift-diffusion model with height
dependent effective temperature.
We show that pullers develop a distal region of constant density
(a supernatant) whose emergence depends on both the activity/gravity
ratio and on the confining geometry (i.e. the cell aspect-ratio).
We also address the statistics of the orientation of bacterial
swimming, finding that, in the regime of small tumbling frequency, the
suspension develops a polar order whose characteristics are strongly
dependent on the type of swimmer.

\section{Numerical method and simulation details} 
The velocity field of the solvent (of dynamic viscosity $\eta$) is evolved by means of a lattice Boltzmann (LB) 
method~\cite{Succi} with nineteen lattice speeds in three dimensions (D3Q19)~\cite{Desplat}. 
Swimmers are modelled as solid spherical objects of radius $R$. 
The correct momentum exchange and mass conservation
through the set of boundary links (between grid points in and out the
sphere) representing the particles is implemented according to the bounce-back-on-links scheme~\cite{Ladd1,Ladd2,Nguyen}.
In order to mimic the surface deformations inducing microswimmers'
self-propulsion, we adopt a simplified version~\cite{MatasNavarro} of 
the ``squirmer'' model~\cite{Blake,Ishikawa}, whereby only the first two terms in the
series expansion of the axisymmetric surface slip velocity profile are retained,
thus leaving just two relevant parameters,
dubbed $B_1$ and $B_2$. The first is related to the propulsion speed, which
is $\mathbf{v}_p = \frac{2}{3}B_1\hat{m}$, where $\hat{m}$ is the squirmer orientation unit vector,
defining the instantaneous swimming direction. The second parameter, $B_2$, determines the
strength of the stresslet, $\mathcal{S} \propto \eta R^2 B_2$, generated by the swimmer in the surrounding fluid
(and, hence, it is related to the amplitude of the injected vorticity)~\cite{Ishikawa}.
The ratio $\beta \equiv \frac{B_2}{B_1}$, such that $\beta \in (-\infty,+\infty)$, 
quantifies the relative intensity of apolar stresses and polar
self-propulsion and classifies swimmers in ``pushers'', $B_2 <0$
(including bacteria like, e.g. {\it E. Coli}), ``pullers'', $B_2>0$ (as the alga {\it Chlamydomonas}),
and ``potential'' swimmers, $\beta=0$ (i.e. swimmers that simply self-propel without generating
vorticity, like the alga {\it Volvox carteri} or certain artificial swimmers)~\cite{Ishikawa,Lighthill,Drescher,Thutupalli,Evans}. 
Every $\tau$ time steps the particles randomize their orientation $\hat{m}$,
thus accounting for the characteristic ``run-and-tumble'' mechanism, which can be seen as 
a source of diffusion for particles that, we recall here, are insensitive to thermal fluctuations.
It is worth noticing that our model, featuring finite size resolved particles, equipped
with the squirming motion, is able to capture hydrodynamic effects in the sedimentation of active
suspensions, both in their near and far field manifestations.\\
We simulate suspensions, of volume fraction $\phi=0.07$, in three-dimensional boxes of size $L \times L \times H$, with height 
$H \approx 80 R$ and variable aspect-ratio
$\Gamma = L/H$. Two solid walls (with no-slip boundary condition for the fluid velocity) confine the system  
in the $z$-direction, while periodic boundary conditions along $x,y$ apply. 
The number of particles, with radius $R=2.3$ (in lattice-spacing units), 
range between $\sim 500$ and $\sim 3 \times 10^4$).
We introduce a reference velocity, $v_g = \mu F_g/(6 \pi \eta R)$  
(where $F_g$ is the gravity force magnitude and $\mu = 1/(6\pi \eta R)$ is the particle mobility), 
i.e. the sedimentation velocity of a passive particle,
and a reference time, $t_c = R/v_p$, that is basically 
the time an isolated particle takes to displace its own radius.
In terms of $v_g$ and $t_c$, the following dimensionless parameters can be defined, namely:
\begin{equation}\label{eq:param}
\chi_1 = \frac{v_p}{v_g} = \frac{2 B_1}{3 v_g}; \quad \chi_2 = \frac{v_{B_2}}{v_g}=\frac{B_2}{3v_g}; \quad \bar{\tau} = \frac{\tau}{t_c},
\end{equation}
that, together with $\beta$, govern the squirmers motion.
In order to investigate how the bacterial swimming characteristics and the system geometry affect the sedimentation profiles, 
we performed several runs exploring the parameter space spanned by $(\chi_1,\chi_2,\beta,\Gamma)$ (the tumbling rate
will be fixed to $\bar{\tau}\approx 4.3$, unless differently specified).

\section{Bacterial density profiles} 
\begin{figure}
\begin{center}
  \advance\leftskip-0.55cm
  \includegraphics[scale=0.55]{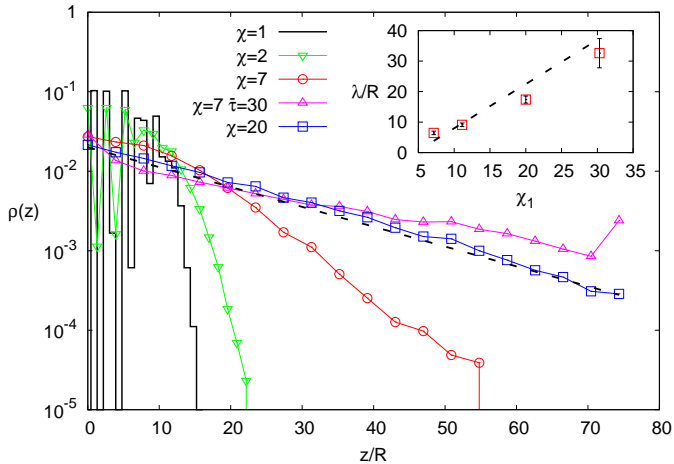}
  \caption{MAIN PANEL: Density profiles $\rho(z)$ for various values of the gravity/propulsion ratio $\chi_1$,
    at $\beta=0$. For $\chi_1$ close to one, bacteria accumulate at the bottom wall, showing
    crystal order (as the regularly spaced peaks in $\rho$ suggest). 
    For large $\chi_1$ the expected exponential profile is recovered. 
    INSET: Dependence of the sedimentation length $\lambda$ (computed out of exponential fits
    of the bacterial density profiles) ($\Box$) on the propulsion/gravity ratio $\chi_1$.  
    The dashed line depicts the theoretical expectation 
    $\lambda/\chi_1 = \ell/3 \approx 1.45 R$, Eq.~(\ref{eq:ltheo}), valid for $\chi_1 \gg 1$.}
\label{fig:densities64}
\end{center}
\end{figure}

\begin{figure}
\begin{center}
  \advance\leftskip-0.55cm 
  \includegraphics[scale=0.55]{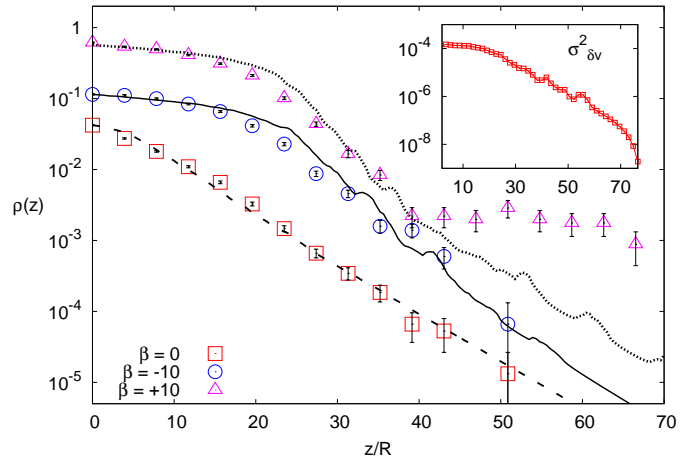}
  \caption{MAIN PANEL: Density profiles for squirmers with $\chi_1 = 10$ and 
   $\beta = 0,\pm 10$ (data are vertically shifted for clarity).
   The lines represent the predictions coming from the
   numerical integration of Eq.~(\ref{eq:model}) with $\lambda = 15$ and $\alpha_1 = 1$ 
  (see the text for the discussion of the 
  model parameters) for $\beta = 0$ (dashed line),  
  $\beta = -10$ (solid line) and $\beta = +10$ (dotted line). 
  INSET: Fluid velocity fluctuations
  $\sigma_{\delta u}^2(z)=\sum_{i=1}^3\langle(u_i(\mathbf{r},t)-\langle u_i(\mathbf{r},t)\rangle)^2 \rangle$
  as a function of the system height for the case $\beta=-10$.}
\label{fig:dens+model}
\end{center}
\end{figure}

We start each run with the bacteria homogeneously distributed in space, with random 
orientations.  
To check that a (non-equilibrium) statistically steady state is reached, we follow the time evolution of the average height  
$h(t) = \frac{1}{H}\int_0^H z \rho(z,t) dz$, 
where $\rho(z,t)$ is the (unsteady) normalized particle density  
(i.e. $\rho(z,t)dz$ is the probability of finding 
a bacterium centred between $z$ and $z + dz$ at the time $t$). We consider as steady state the time interval during
which $h(t)$ fluctuates by less than $\sim 5 \%$. All data shown hereafter are meant averaged over such time interval.
Our aim is to study the impact that activity, in terms of $\chi_1$ and $\beta$, has on the squirmer sedimentation,
and to characterize the emerging dynamical regimes,
checking whether and how hydrodynamic effects come into play. 
According to the theory~\cite{Tailleur1,Tailleur2},
as $\chi_1 \rightarrow 1$, all particles concentrate at the bottom wall.
Instead, when $\chi_1 \gg 1$ (i.e., in the self-propulsion dominated regime)
the steady state sedimentation profile should display an exponential form
$\rho (z) \sim e^{-z/\lambda}$,  
with a sedimentation length depending on the single particle velocity
(and, hence, on $\chi_1$) as 
\begin{equation}\label{eq:ltheo}
\lambda = \frac{v_p^2 \tau}{3 v_g} = \frac{\ell}{3}\chi_1,
\end{equation}
where $\ell = v_p \tau$ is the bacterium run length.
This result has been found in agreement with experimental observations~\cite{Palacci} and
numerical simulations~\cite{Nash}.
The exponential profile characterizes also equilibrium systems,
as in the classical Perrin's experiment for (thermal) colloids~\cite{Perrin};
the sedimentation length is determined by the particle diffusivity, $D$, and the
gravity force as $\lambda^{(eq)} = D/(\mu F_g)$ and depends, therefore, through the Stokes-Einstein relation
$D=\mu k_B T$, on the system temperature $T$, namely $\lambda^{(eq)}=k_B T/F_g$. The formal analogy with
the passive (equilibrium) case suggests, then, to introduce an {\it effective} temperature as:
\begin{equation}\label{eq:T1p}
k_B T_{\mbox{\tiny{eff}}}^{(1p)} = \frac{v_p^2 \tau}{3 \mu},
\end{equation}
such that the sedimentation length reads $\lambda=k_B T_{\mbox{\tiny{eff}}}^{(1p)}/F_g$. 
In Fig.~\ref{fig:densities64} we plot the time-averaged steady state bacterial density profiles
for various values of $\chi_1$ and $\beta=0$.
For values close to one, as expected, bacteria uniformly fall down under the action of gravity;
however, due to the particles finite size, the sedimentation
length remains finite.
The particles in the sediment tend to organize themselves in layers with crystal-like order,
noticeable from the peaks in the density profile, close to the bottom
wall, displaced from each other by about one diameter ($2R$), as found also in a previous computational study~\cite{Kuhr}. 
At increasing $\chi_1$, swimmers occupy an increasingly larger volume of liquid, and correspondigly 
$\rho(z)$ shows, over the whole box length, the predicted exponential profile~\cite{Tailleur2}, with 
a sedimentation length growing linearly with $\chi_1$ (see inset of Fig.~\ref{fig:densities64}).\\
If we increase $|\beta|$ (thus intensifing bacterial activity) to large
enough values, for a fixed $\chi_1$, the deviation
from the exponential profile can be important, as one can see from Fig.~\ref{fig:dens+model},
where we plot the bacterial density $\rho(z)$ for 
three cases with same $\chi_1 = 10$ but with different $\beta$. 
In the pushers/pullers case ($\beta \neq 0$), dynamic correlations are so intense 
that recovering a Perrin-like form just with the introduction of
a global effective diffusion coefficient as coming from {\it single} particle
is no longer possible~\cite{Palacci}.
The larger $|\beta|$, the stronger is the departure of the 
sedimentation profile from an exponential; indeed we found
that deviations start to be relevant from $|\beta| \approx 5$ on. 
This observation may justify why in a previous numerical study of
sedimenting bacteria with hydrodynamic interactions~\cite{Nash} (whose
force dipole model would give an equivalent $\beta = -1$) apparently
no significant effects were detected. 

\section{Extended diffusive model} 
Due to hydrodynamic correlations the dynamics of a bacterium in the 
suspension is affected by the presence of the others through the generation of motion within
the liquid, which will act as a ``bath'' at an effective temperature
(that measures the fluid ``agitation'').
We can understand these effects extending a diffusive model proposed earlier to
describe bacterial
sedimentation~\cite{Palacci},
based on the Smoluchowski equation $\partial_t \rho = - \nabla \cdot \mathbf{J}$,
determined by the flux $\mathbf{J} = -\tilde{D} \mathbf{\nabla} \rho + \tilde{\mu} \mathbf{F}_g \rho$.
The ratio of the {\it local} diffusion coefficient, $\tilde{D}$, and particle mobility, $\tilde{\mu}$,
by virtue of a generalized Stokes-Einstein relation, represents the effective temperature field.
Assuming that in the steady state the
density will only depend on $z$ (we will come back later to the
validity of this assumption), the zero flux boundary conditions at the walls gives:
\begin{equation} \label{eq:ODE}
\frac{d \rho}{d z} = -\frac{F_g}{k_B T_{\mbox{\tiny{eff}}}}\rho.
\end{equation}
We propose an effective temperature of the form
$T_{\mbox{\tiny{eff}}}=T_{\mbox{\tiny{eff}}}^{(1p)}+T_{\mbox{\tiny{eff}}}^{\mbox{\tiny{(coll)}}}$,
consisting of two terms:
the single-particle effective temperature, Eq.~(\ref{eq:T1p}), accounting for the self-propulsion,
plus a contribution proportional to fluid velocity fluctuations, $T_{\mbox{\tiny{eff}}}^{\mbox{\tiny{(coll)}}}$,
capturing the collective effects due to hydrodynamic interactions.
However, since in the steady state bacteria are distributed inhomogeneously over the volume 
(with a density increasing from top to bottom), also fluid velocity fluctuations 
$\sigma_{\delta u}^2 = \sum_{i=1}^3\langle (u_i(\mathbf{r},t) - \langle u_i(\mathbf{r},t) \rangle)^2 \rangle$
(where $\langle (\cdots) \rangle = \frac{1}{L^2}\int \int (\cdots) dx dy$) are expected 
to vary (as indeed it can be seen in the inset of Fig.~\ref{fig:dens+model}). 
We should then cope with a height dependent effective temperature 
$T_{\mbox{\tiny{eff}}}(z)=T_{\mbox{\tiny{eff}}}^{(1p)}+T_{\mbox{\tiny{eff}}}^{\mbox{\tiny{(coll)}}}(z)$, 
leading, upon insertion in (\ref{eq:ODE}), to an equation for the sedimentation density which can 
be recast in the following form
\begin{equation}
\frac{d \rho}{d z} = -\frac{1}{\lambda}
\frac{\rho}{\left(1+\frac{T_{\mbox{\tiny{eff}}}^{\mbox{\tiny{(coll)}}}(z)}{T_{\mbox{\tiny{eff}}}^{(1p)}}\right)}, 
\end{equation}
where $\lambda = (k_B T_{\mbox{\tiny{eff}}}^{(1p)})/F_g$ is the  sedimentation length discussed 
in the previous section. 
We assume, then, $T_{\mbox{\tiny{eff}}}^{\mbox{\tiny{(coll)}}}(z) \propto \sigma_{\delta u}^2(z)$ to hold, 
so that we can finally write
\begin{equation} \label{eq:model}
\frac{d \rho}{d z} = -\frac{1}{\lambda}\frac{\rho}{\left(1+\alpha_1 \frac{\sigma_{\delta u}^2 (z)}{v_p^2}\right)},
\end{equation}
with $\alpha_1$ a free parameter representing the proportionality 
constant between $T_{\mbox{\tiny{eff}}}^{\mbox{\tiny{(coll)}}}$ and $\sigma_{\delta u}^2$.
Comparing the numerical integration of Eq.~(\ref{eq:model}) with data from LB simulations
(see Fig.~\ref{fig:dens+model}), we find that the proposal of gauging the global effective temperature to a height dependence 
works well for $\beta = 0$ and $\beta < 0$.
The phenomenology of pullers ($\beta > 10$) appears, however, to be more complicated: 
in fact, while the density 
profile can be recovered where the concentration is higher, the presence of 
a region of constant density, denoting the formation of a supernatant floating over the sedimentated
layer, eludes the generalized diffusive model. 

\section{The case of ``shakers''}  
\begin{figure}
\begin{center}
  \advance\leftskip-0.55cm 
  \includegraphics[scale=0.35]{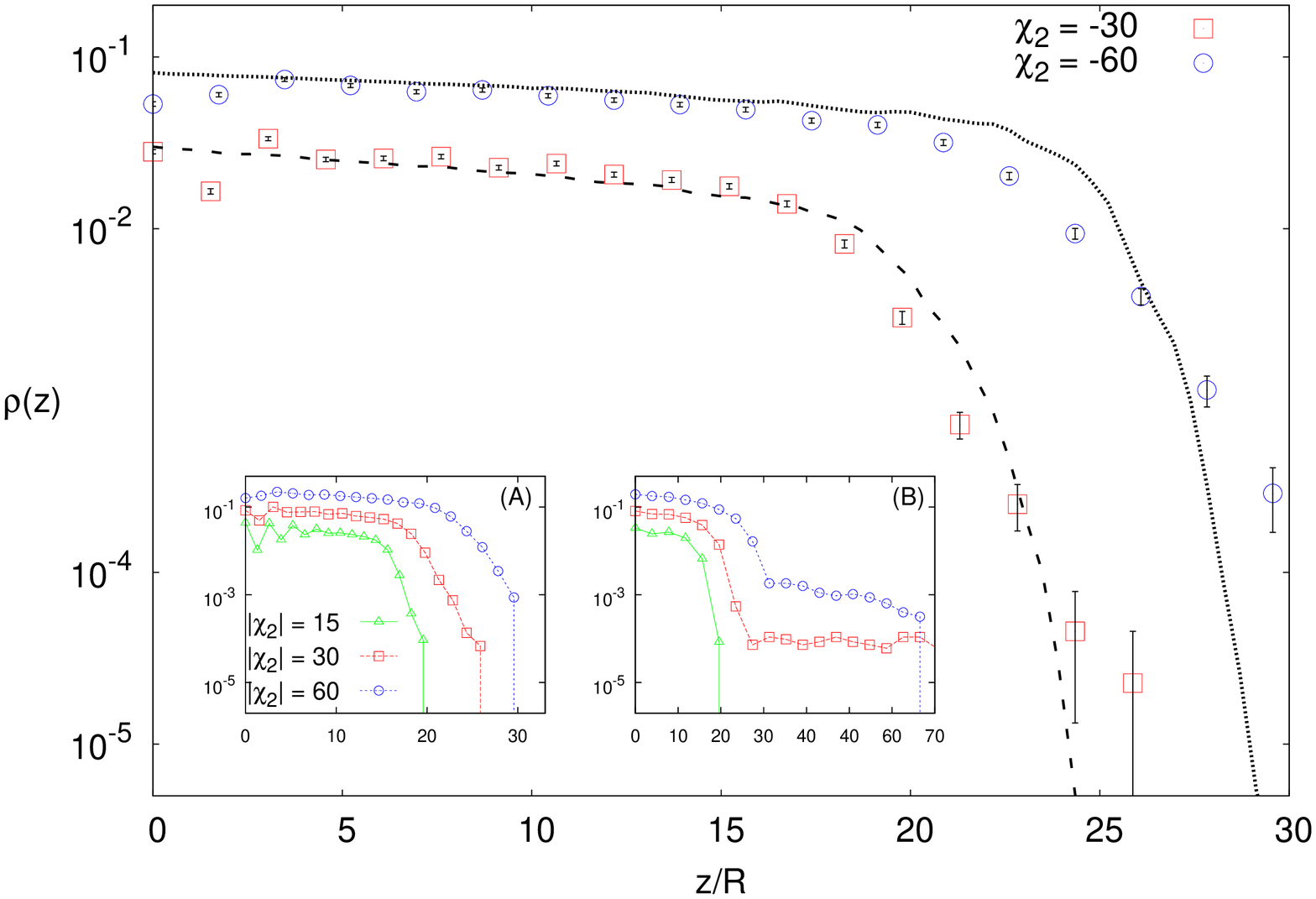}
  \caption{MAIN PANEL: Density profiles for shakers with two different $\chi_2<0$ (here and in the insets data
   are vertically shifted for clarity): the larger $|\chi_2|$ the longer the density tail (i.e. the 
   wider is the region occupied by particles). 
   The lines are the theoretical predictions coming from the numerical integration 
   of Eq.~(\ref{eq:model2}), where the function $\sigma_{\delta u}^2(z)$ is taken from the simulations,
   with $\alpha_2=4.4$.
   INSET A: Density profiles for shakers with negative $\chi_2$.
   INSET B: Density profiles for shakers with positive $\chi_2$: notice the 
   formation of the supernatant for large enough $\chi_2$.} 
\label{fig:denshakers+theory}
\end{center}
\end{figure}

\begin{figure}
\begin{center}
  \advance\leftskip-0.55cm 
  \includegraphics[scale=0.55]{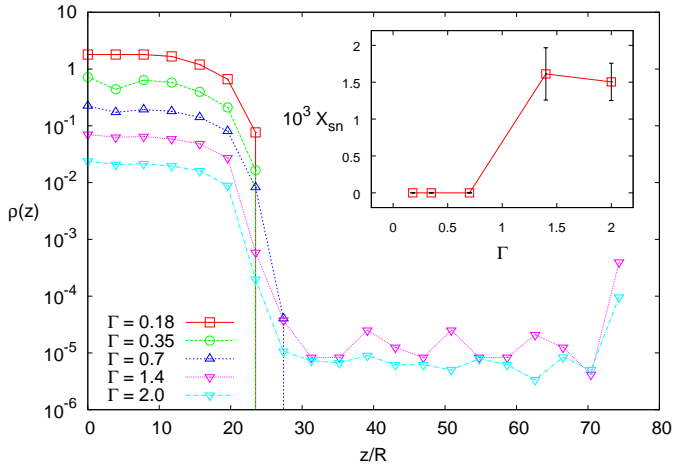}
  \caption{MAIN PANEL: Density profiles for shakers with $\chi_2 = 8.3$ for
    various aspect-ratios $\Gamma = L/H$ (data are vertically shifted for clarity). INSET: Fraction of particles
    in the supernatant region, computed as $X_{\text{sn}}=\int_{\zeta_0}^H \rho(z)dz$ ($\zeta_0$ being the minimum height such that $\rho(z)=0$, for some $\Gamma$ and for $z>\zeta_0$),
    as function of the aspect-ratio of the
    cell: notice that for $\Gamma < 1$, $X_{\text{sn}}=0$, i.e. no supernatant develops.} 
\label{fig:ARsupernat}
\end{center}
\end{figure}

Another striking instance of how crucial the role played by hydrodynamics can be is provided by the
regime where 
$|\beta| \rightarrow \infty$, i.e. $B_1$ goes to zero while $B_2$ stays finite.
This regime corresponds to active suspensions where 
particles do not self-propell but generate motion in the fluid and are relevant for microswimmers
known as ``shakers''~\cite{Marchetti,Ramachandran}, like, e.g., melanocytes~\cite{Gruler}.  
Since both their propelling velocity and the effect of thermal fluctuations are negligible,
such a suspension would undergo a gravitational collapse,
if one could completely neglect the presence of the solvent.
However, as shown in Fig.~\ref{fig:denshakers+theory}, 
the steady state density profiles develop a sedimentation
layer, whose width increases with $\chi_2$. 
The observed width cannot be interpreted simply as a result of the
close-packing of the particles, which would imply, in fact, a value of around $8R$,
much smaller than the measured one.
We try to recover the sedimentation profiles
of shakers following the same ideas of the previous section. We must integrate numerically 
an analogue of Eq.~(\ref{eq:model}), the main difference being that now the one-particle contribution 
to the effective temperature $T_{\mbox{\tiny{eff}}}^{(1p)}$ is zero, since an isolated shaker does not self-propell,
so that we get 
$T_{\mbox{\tiny{eff}}}(z)=T_{\mbox{\tiny{eff}}}^{\mbox{\tiny{(coll)}}}(z) = \alpha_2(\sigma_{\delta u}^2 (z)/v_{B_2}^2)$ 
(here we indicate the phenomenological parameter as $\alpha_2$ in order to distinguish
it from that of propellers).
The reference speed $v_{B_2}=B_2/3$, implicitly introduced in (\ref{eq:param}), is the magnitude of the
velocity field generated by an isolated shaker, averaged over its surface.
The stationary Smoluchowski equation, then, reads:
\begin{equation} \label{eq:model2}
\frac{d \rho}{d z} = -\frac{F_g}{\alpha_2}\frac{\rho}{\left(\sigma_{\delta u}^2 (z)/v_{B_2}^2 \right)};
\end{equation}
the results of the numerical integration of Eq.~(\ref{eq:model2}) for shakers with negative $\chi_2$ with 
two different values of gravity are reported in Fig.~\ref{fig:denshakers+theory}, showing, again, good agreement.
Analogously to the case of pullers, shakers with $\chi_2>0$ develop (for $\chi_2$
large enough) a distal region of constant density in the sedimentation
profile (see inset B of Fig.~\ref{fig:denshakers+theory}).
The emergence of such supernatant is due to the sedimentate which
acts as a pump and generates motion in 
higher layers of fluid. It is, then, a genuinely
{\it three-dimensional}  and {\it non-local} effect, two
features which make also our formalism based on a height
dependent effective temperature fail.
To support this picture, we show that, for a fixed value of $\chi_2$
the supernatant disappears when decreasing the aspect-ratio $\Gamma$
of the cell below unity (see Fig.~\ref{fig:ARsupernat}). 
This is, indeed, a
manifestation of three-dimensionality: 
using an analogy with a Rayleigh-B\'enard system~\cite{Chandrasekhar}, we
argue that the geometry favors (or not) the 
development of a large scale flow which can (or can not) sustain the supernatant.
In fact, the difference in the fluid flow pattern generated by a single particle,
either a pusher or a puller, is not strong enough to sustain the different macroscopic
patterns observed if the swimmers are randomly oriented (as a matter of fact, no supernatant is observed
for pushers, or shakers, with $\beta<0$).
Hence, a collective organization of the swimmers is required to produce the observed macroscopic flows.
We will next address  the emergence of orientational order in the sedimentaing profiles of microswimmers.

\section{Orientational statistics}
\begin{figure}
\begin{center}
  \advance\leftskip-0.55cm
  \includegraphics[scale=0.45]{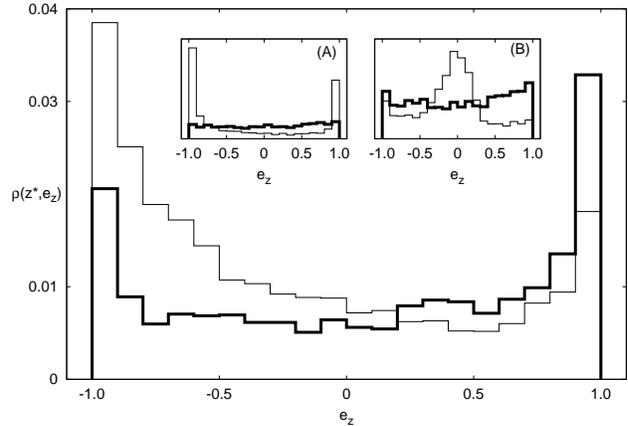}
  \caption{Probability distributions of the vertical component of bacterial orientation, $e_z$, measured inside two slabs $[z^{\ast}-2R, z^{\ast}+2R]$,
    with $z^{\ast} = 2R$ (thin line) and $z^{\ast} = 18R$ (thick line), respectively.
   MAIN PANEL: Potential swimmers ($\beta =0$). INSET A:  
   Pullers ($\beta > 0$). INSET B: Pushers ($\beta < 0$). 
   In all cases $\chi_1 = 10$ and $\bar{\tau} \approx 30$.}
\label{fig:JPDF}
\end{center}
\end{figure}

The emergence and the dynamical relevance of anisotropic ordering in active fluid systems has been 
widely recognized in the literature~\cite{Ramaswamy,Marchetti,Enculescu}. 
We study the orientational statistics measuring the PDF $\rho(z^{\ast},e_z)$ of the $z$-component of the bacterial {\it squirming} characteristic vector, $e_z$, within slabs of width $4R$ centred at 
different heights $z^{\ast}$ along the cell.
For squirmers  with $\chi_1 = 10$, $\beta=0$ and tumbling time $\bar{\tau} \approx 4.3$ 
we find a bimodal distribution symmetrically peaked at $e_z = \pm 1$
(with a slight imbalance towards $e_z = -1$) and almost insensitive to changes in $z^{\ast}$.
However, for $\bar{\tau} \approx 30$, we observe that, close to the wall, the peak in $\rho(z^{\ast}=2R,-1)$
is more pronounced than that in $\rho(z^{\ast}=2R,+1)$, whereas the opposite trend appears at  
$z^{\ast}=18R$ (see Fig.~\ref{fig:JPDF}), which means that in the bulk bacteria swim preferentially upwards
(i.e. against gravity).
Previous theoretical studies have predicted the emergence of polar order
when bacterial self-propulsion dominates over thermal noise~\cite{Enculescu}.
In our athermal case, where tumbling plays the role of an effective noise,  
the higher the tumbling rate (short $\bar{\tau}$) the closer is the dynamics to thermal diffusion. 
Therefore, we need to increase $\bar{\tau}$
in order to favour the {\it active} diffusion due to self-propulsion and, consequently, the suspension polarization.
Analogously, for $\beta \neq 0$ we expect this scenario to break down,
because the generation of fluid motion acts as an effective source of ``noise''; 
indeed, we observe that, close to the wall, $\rho(z^{\ast},e_z)$ is peaked around $e_z \approx 0$
for $\beta < 0$, and it is bimodal (with a higher at $e_z \approx -1$) for $\beta > 0$,
while in the bulk it is rather uniform in both cases (insets A and B of Fig.~\ref{fig:JPDF}).
As anticipated in the previous section, such different orientational
ordering between pushers and pullers turns out to have an impact also
on the swimmers' distribution in space, as indicated by the
sedimentation profiles. 

\section{Conclusions} 
We have presented a computational study of suspensions of run-and-tumble squirmers under gravity.
Thanks to the built-in properties of the mesoscopic approach adopted we could 
take into account both the finite size of particles and the hydrodynamics of the solvent.
In the case of potential swimmers, agreement has been found with 
theoretical predictions regarding i) the dependence of the density profiles on the activity/gravity ratio and
ii) the emergence of a polar order from the inspection of distributions of particle orientations.
We have reported evidence that, for pushers and pullers with large enough $\beta$, 
the hydrodynamic flows induced by their collective motion 
determine sedimentation profiles that cannot be understood in terms of single swimmer response to the
gravitational field. This observation appeared particularly distinctive in the emblematic case of shakers.
We have, therefore, generalized the theory
on the basis of a height dependent
{\it collective} effective temperature. 
Moreover, we have provided instances of cases (i.e. pullers and
shakers with positive $\beta$) where, due to the fluid
flow and to a non-trivial organization of swimmers, 
the full {\it three-dimensional} dynamics must be considered for the sake
of a satisfactory understanding of the sedimentation phenomenology.

\section*{Acknowledgements}
We acknowledge MICINN and DURSI for financial support under
Projects No. PGC2018-098373-B-I00, and No. 2017SGR-
884, respectively. IP acknowledges SNSF for financial support under
Project No. 200021-175719.
This work was possible thanks to the access to MareNostrum
Supercomputer at Barcelona Supercomputing Center (BSC) and
also through the Partnership for Advanced Computing in Europe
(PRACE).

\bibliography{bacteria-sedimentation_arxiv} 

\end{document}